\documentclass{iau} 
\usepackage{graphicx}

\newcommand{\mnras}{\textit{MNRAS}}
\newcommand{\apj}{\textit{ApJ}}
\newcommand{\aj}{\textit{AJ}}
\newcommand{\apjl}{\textit{ApJ (Letters)}}
\newcommand{\apjs}{\textit{ApJ (Suppl.)}}

\newcommand{\nat}{\textit{Nature}}
\newcommand{\araa}{\textit{ARAA}}

\title[Atomic and Molecular Phases of the ISM]{Atomic and Molecular Phases of the Interstellar Medium}

\author[Mordecai-Mark Mac Low]   
{Mordecai-Mark Mac Low$^1$}

\affiliation{$^1$Dept.\ of Astrophysics, American Museum of Natural
  History, Central Park West at 79th Street, New York, NY 10024-5192,
  USA and Institut f\"ur Theoretische Astrophysik, Zentrum f\"ur
  Astronomie der Universit\"at Heidelberg, Albert-Ueberle-Str.\ 2,
  69121 Heidelberg, Germany \\ email:{\tt mordecai@amnh.org}} 

\pubyear{2016}
\volume{315}  
\jname{From interstellar clouds to star-forming
  galaxies: universal processes?}
\editors{F. Van der Tak, P. Jablonka, \& P. Andr\'e, eds.}
\begin{document}

\maketitle

\begin{abstract}
  This review covers four current questions in the behavior of the
  atomic and molecular interstellar medium. These include whether the
  atomic gas originates primarily in cold streams or hot flows onto
  galaxies; what the filling factor of cold gas actually is in
  galactic regions observationally determined to be completely
  molecular; whether molecular hydrogen determines or merely traces
  star formation; and whether gravity or turbulence drives the
  dynamical motions observed in interstellar clouds, with implications
  on their star formation properties.

\end{abstract}

\firstsection 
\section{Introduction}

Star formation occurs primarily in atomic and molecular interstellar
gas, although \cite[Keto (2002)]{Keto02} emphasizes that around a massive
enough star, even ionized gas can be accreted. In this review, I
discuss a number of recent advances in our understanding of the
origin and dynamics of that gas and its relation to star formation.  

\section{Cold Accretion and its Limits}

\begin{figure}[bth]
\begin{center}
\includegraphics[width=1.8in,angle=90]{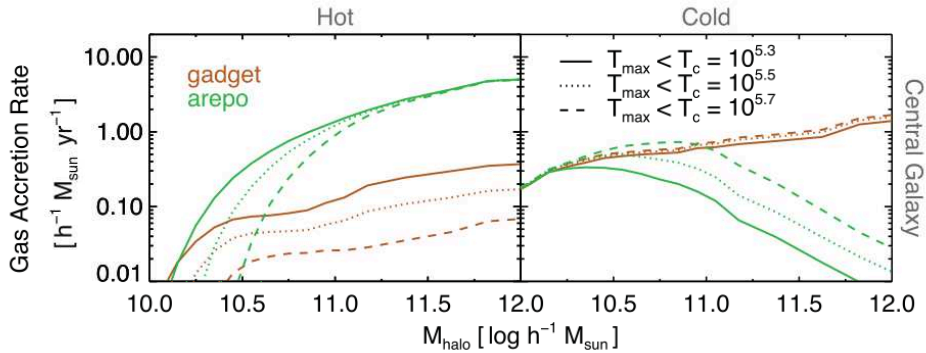} 
\caption{\label{Nelson13} Comparison of gas accretion rates for hot
  gas (left, with temperature above $T_c$ as shown in legend) and cold
  gas (right) as a function of halo mass, for SPH models (brown) and
  Voronoi mesh models (green).  The two methods agree for small masses
  typical of high redshifts, but the Voronoi mesh code predicts most
  gas will reach high temperatures for large halo sizes, disagreeing
  with the SPH results. Adapted from \cite[Nelson et al.\ (2013)]{Nelson13}.}
\end{center}
\end{figure}

The origin of this
gas lies in gas accreted onto galaxies during their formation and
subsequent evolution.
Perspectives on how that gas arrives have changed several times
recently.  Originally, it was thought that all gas accreted into the
halo of a galaxy the size of the Milky Way would be heated to the
virial temperature of over $10^6$~K, followed by gradual cooling and accretion.
However, a decade ago, numerical simulations started to suggest that a major,
possibly even dominant alternative was for dense
streams to accrete from the cosmic web, dense enough to cool quickly
when compressed and thus to never heat to the virial temperature
(e.g. \cite[Fardal et al.\ 2001]{Fardal01}, \cite[Birnboim \& Dekel
2003]{Birnboim03}, \cite[Kere\v{s} et al.\ 2005]{Keres05},
\cite[2009]{Keres09}, \cite[Ocvirk et al.\ 2008]{Ocvirk08},
\cite[Dekel \& Birnboim 2006]{Dekel06}, \cite[Dekel et al.\ 2009]{Dekel09},
\cite[2013]{Dekel13}, \cite[Goerdt et al.\ 2015]{Goerdt15}). However,
\cite[Nelson et al.\ (2013)]{Nelson13}, using the AREPO Voronoi mesh code
(\cite[Springel 2010]{Springel10}), argued that cold streams are
likely only important for small galaxies at high redshift (see Fig.~\ref{Nelson13}), while their
appearance at lower redshifts and for larger galaxies is a numerical
artifact of the smoothed particle hydrodynamics (SPH) method, as exemplified
in that paper by GADGET (\cite[Springel 2005]{Springel05}).

Evidence for continuing accretion of gas in the modern Milky Way can
be gained from observations of high velocity clouds of neutral gas
falling towards the galactic plane at velocities over 70 km~s$^{-1}$.
Material falling in at slower velocities has been shown to have
Galactic chemical abundances, while the high velocity clouds have far
lower abundances, suggesting extragalactic origin, as reviewed by \cite[Putman et al.\
(2012)]{Putman12}. Simple analytic models modeled on the flow of water into a bathtub and
down the drain can guide our understanding of the cycle of gas into the
atomic phase from infalling ionized gas, and then out of it again
through star formation or expulsion of the gas in galactic outflows
(\cite[Bouch\'e et al.\ 2010]{Bouche10}, \cite[Dekel et al.\
2013]{Dekel13}, \cite[Dekel \& Mandelker 2014]{Dekel14}). 

\section{Why a Cold, Dense, ISM is Mostly Hot}

\begin{figure}[bth]
\begin{center}
\includegraphics[width=2.5in,angle=90]{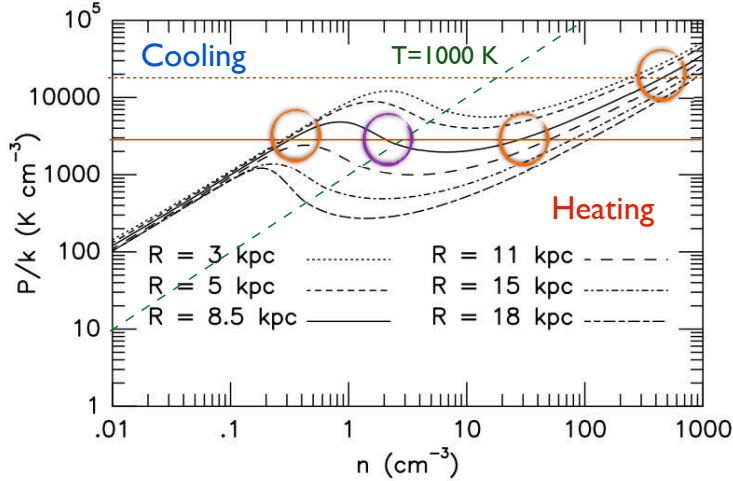} 
\caption{\label{Wolfire03} Thermal equilibrium curves at different
  radii in the Milky Way, compared to a typical pressure near the
  Solar circle (solid orange line), and compared to the pressure in the molecular ring or
  starburst galaxies (dashed orange line).  Orange circled points represent
  stable equilibria, where small perturbations of temperature (motions
  perpendicular to the green-dashed line, with increasing
  temperature towards the left) result in restoring
  increases in cooling or heating, as labeled. Purple circled point is
  unstable. Adapted from \cite[Wolfire et al.\ (2003)]{Wolfire03}.}
\end{center}
\end{figure}

The thermodynamic state of the atomic and diffuse molecular gas is controlled
by the balance between optically-thin radiative cooling and photoelectric heating by far ultraviolet
radiation.  The between around 100~K and $10^4$~K the
cooling curve is a sufficiently steep function of temperature to
produce a region of thermal instability (\cite[Field et al.\
1969]{Field69}, \cite[Wolfire et al.\ 1995]{Wolfire95},
\cite[2003]{Wolfire03}). In the case of pressures typical of the Milky
Way, this leads to a two-phase medium (solid line in
Fig.~\ref{Wolfire03}).  When pressures increase (dotted line), only
the cold, dense phase remains thermally stable.

\cite[Ostriker et al.\ (2010)]{Ostriker10}
argued that, in fact, this transition leads to self-regulation of star
formation. The pressure range within which a warm phase will be present is
determined by the far ultraviolet radiation field produced by stars.
If the vertical hydrostatic pressure exceeds that range, it will drive
nearly all the gas into the cold phase, promoting star formation, and
thus increasing the far ultraviolet field until a warm medium can be
reestablished. Thus, the combination of thermal and hydrostatic
equilibrium controls,  in the terms of the bathtub model, how
fast the bathtub drains.

\begin{figure}[bth]
\begin{center}
\includegraphics[width=2in,angle=90]{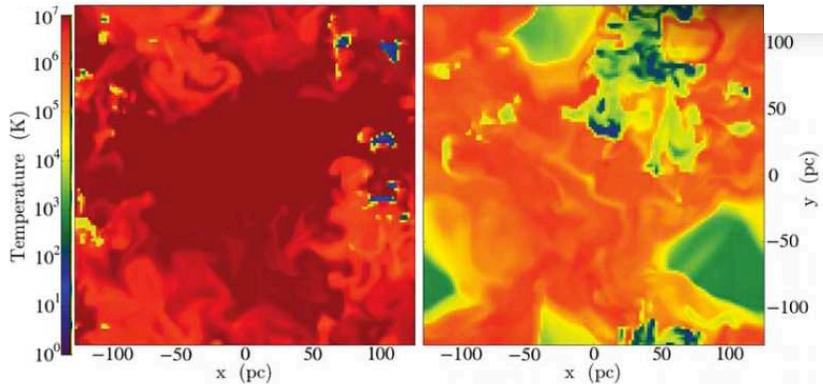} 
\caption{\label{Gatto15} Temperature distributions on slices through
  periodic boxes of supernova-driven turbulence.  On the left, supernovae
  explode in random positions, while on the right, half the supernovae
  explode in density peaks, substantially increasing the loss of
  energy to radiative cooling.  On the left, pressures have reached
  high enough levels to eliminate the warm stable phase. Adapted from
  \cite[Gatto et al.\ (2015)]{Gatto15}.} 
\end{center}
\end{figure}

While the point that high enough pressures produce a predominantly
cold temperature distribution is widely understood, what is perhaps less appreciated
is the resulting spatial distribution of the gas.  \cite[Gatto et
al.\ (2015)]{Gatto15} and \cite[Li et al.\ (2015)]{MiaoLi15} emphasize
that the filling factor of the cold, dense gas will be small in such a
high-pressure environment.  In regions with supernova explosions
(either from active star formation or even Type Ia supernovae from
older populations) the space in between the cold, dense clumps is
filled with hot, tenuous gas (see Fig.~\ref{Gatto15}), the third phase of \cite[McKee \&
Ostriker (1977)]{McKee77}. 

Observational tracers of atomic and molecular gas will show that the
mass of the interstellar gas is almost entirely in cold, molecular
material, leading to the description of the region as a fully
molecular interstellar medium.  The hot gas, on the other hand, is
difficult to observe, as its density is low and temperature high
enough that it emits primarily in bremsstrahlung. However, it
still completely dominates the volume filling factor, and needs to be taken into
account in any model of high-pressure, fully molecular, interstellar
gas.  \cite[Joung et al.\ (2009)]{Joung09} demonstrated that following
the Kennicutt-Schmidt law for star formation in stratified disks
results in a three-phase medium, while \cite[Walch et al.\
(2015)]{Walch15} showed that increasing the star formation rate by a
factor of three already leads to a cold interstellar medium filled
with hot gas.

 \section{Molecules are a Consequence of Star Formation}

Observations have demonstrated a linear correlation between the
surface density of star formation and that of molecular hydrogen
(\cite[Rownd \& Young 1999]{Rownd99}, \cite[Wong \& Blitz
2002]{Wong02}, \cite[Bigiel et al.\ 2008]{Bigiel08}, \cite[Leroy et
  al.\ 2008]{Leroy08}).  This raises the question of whether this
  correlation implies a causal relationship.  Is molecular hydrogen
  formation required for star formation to occur?

It has more recently become evident that many tracers of high density
gas correlate just as well with the star formation rate as does
molecular hydrogen.  For example, the tracer molecule HCN was shown by
\cite[Gao \& Solomon (2004)]{Gao04} to also correlate with star
formation rate in both normal spiral galaxies and luminous and
ultraluminous infrared galaxies.  In the solar neighborhood, the
number of young stellar objects in a cloud correlates with the mass of
the cloud contained in regions with a K-band infrared extinction $A_K
> 0.8$ (\cite[Lada et al.\ 2010]{Lada10}).  \cite[Clark \& Glover
  2014]{Clark14} indeed found thresholds for star formation
in numerical simulations that reproduce this observed
threshold. However, the physical process producing this threshold is
{\em not} molecule formation, but rather dust shielding from heating
by the background far ultraviolet radiation.

\begin{figure}[bth]
\begin{center}
\includegraphics[width=4in,angle=90]{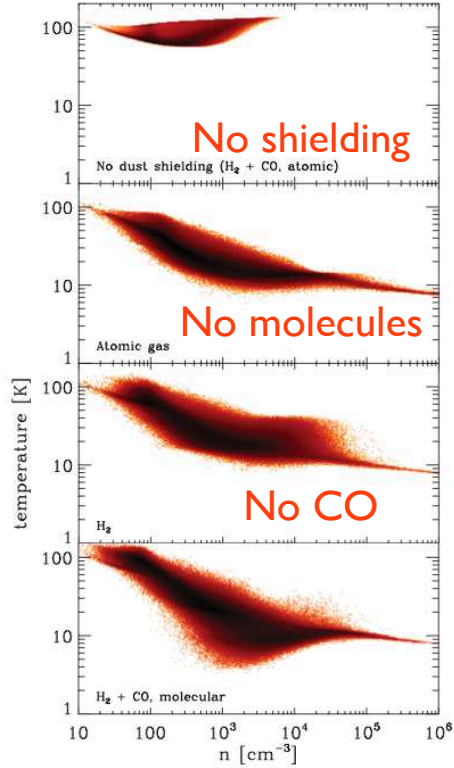} 
\caption{\label{Glover12} Distribution of temperature and pressure in
  models of turbulent, atomic gas including both atomic and molecular
  chemistry and cooling processes, as well as photoelectric heating
  modulated by dust shielding.  The bottom panel shows the full model,
  while the upper panels have successively less physics. The first
  panel from the bottom neglects molecular line cooling, the second
  also neglects molecular hydrogent formation, and the top also
  neglects dust shielding.  Gas reaches 10~K and high densities in all
  but the last of these models. Adapted from \cite[Glover \& Clark (2012)]{Glover12}.} 
\end{center}
\end{figure}

Indeed \cite[Krumholz et al.\ (2011)]{Krumholz11} and \cite[Glover \&
Clark (2012)]{Glover12} have demonstrated that atomic line cooling is
as effective as molecular cooling in bringing gas down to star forming
temperatures of 10-20~K.  Figure~\ref{Glover12} demonstrates that the same
cold, star-forming gas appears in numerical simulations with full
molecular and atomic cooling; with only atomic cooling, but still including
molecular hydrogen formation; and with all molecular physics suppressed, but still
including dust shielding from far ultraviolet heating.  Only the
removal of dust shielding from the simulation suppresses star
formation. 

Thus, it appears that the formation of molecular hydrogen does not control
star formation. Rather, it seems, star formation, or more generally
gravitational collapse of atomic gas, controls molecular hydrogen
formation.

\section{Molecular Cloud Dynamics Driven by Gravity Rather Than
  Turbulence}

Molecular cloud dynamics offer a useful window to their origins.  The
argument has been made that they form by the sweeping up of smaller
clouds in superbubbles or spiral arms (e.g.\ \cite[Tasker \& Tan
2009]{Tasker09}). An alternative proposal is that clouds form from
large scale gravitational collapse, as proposed by
\cite[Hartmann et al.\ (2001)]{Hartmann01} to explain the apparent
correlation of star formation across entire molecular clouds.

\begin{figure}[bth]
\begin{center}
\includegraphics[width=3.5in,angle=90]{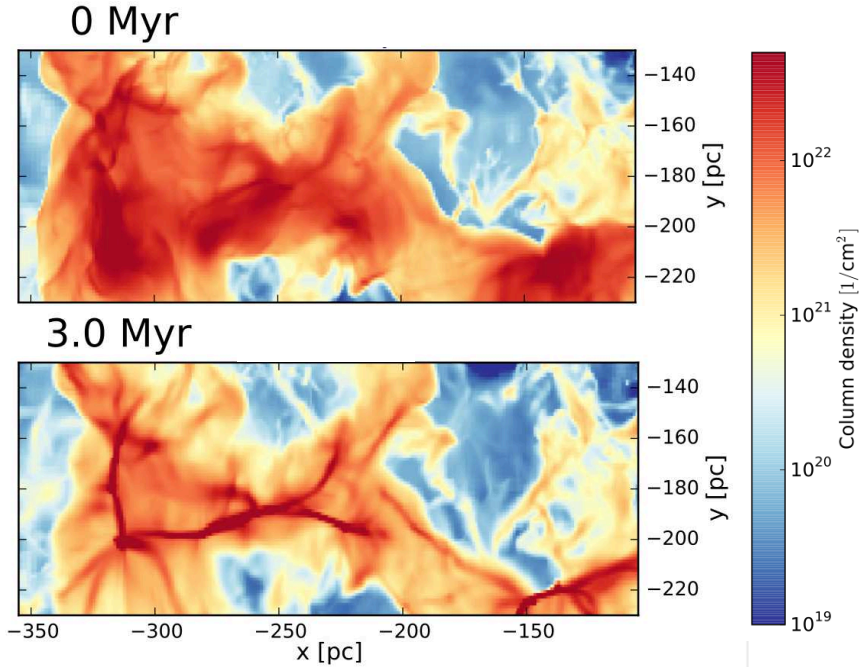} 
\caption{\label{IbanezMejia15} Vertical projection through a
  sub-region of a supernova-driven, stratified, magnetized model of
  the interstellar medium at times of 0 and 3 Myr after self-gravity
  has been turned on. (The full domain is a $1\times 1\times 20$~kpc
  vertical cut through the galactic disk.)  Prompt collapse occurs,
  forming filaments strongly resembling observed IRDCs, with
  star-forming cores strung along them. Adapted from
  \cite[Ib\'a\~nez-Mej\'{\i}a et al.\ (2015)]{IbanezMejia15}.} 
\end{center}
\end{figure}

Models of the supernova-driven, stratified interstellar medium can be
used to examine these scenarios.  One example of such a model derives
from work by \cite[Joung \& Mac Low (2005)]{Joung05}, \cite[Joung et
al.\ (2009)]{Joung09}, and \cite[Hill et al.\ (2012)]{Hill12}.  It
includes both clustered and random supernovae, a background
gravitational potential from the stellar disk and the halo,
photoelectric heating and equilibrium ionization cooling, and magnetic
fields.  These models swept up gas into filaments that appeared
similar to large molecular clouds.  An example of such a filament is
shown in top panel of Figure~\ref{IbanezMejia15}.

\cite[Ib\'a\~nez-Mej\'{\i}a et al.\ (2015)]{IbanezMejia15}
added the self-gravity of the gas to these models, turning it on after
a full vertical galactic fountain had been set up over the course of
over 200~Myr. This long preparation period also allowed plenty of time
for the dense regions to come into equilibrium with the
supernova-driven turbulence in the surrounding diffuse medium.
However, when the self-gravity was turned on, immediate collapse
ensued, as is shown in the example of Figure~\ref{IbanezMejia15}.  One
immediate and notable result was the
enhancement of the filamentary morphology to the extreme levels
characteristic of observed infrared dark clouds (e.g. \cite[Simon et
al.\ (2006)]{Simon06}). 

Within 5 Myr, this collapse lead to the
formation of unresolved dense cores within which rapid star
formation appears inevitable.  As these first models have no mechanism
to simulate the local prompt feedback in ionizing radiation,
stellar winds and so forth from this star formation, we can no longer
follow its evolution after this point.  However it is clear already
that without this local feedback, star formation would reach
unrealistic rates, just as first argued by \cite[Zuckerman \& Palmer
(1974)]{Zuckerman74}.

\begin{figure}[bth]
\begin{center}
\includegraphics[width=1.45in,angle=90]{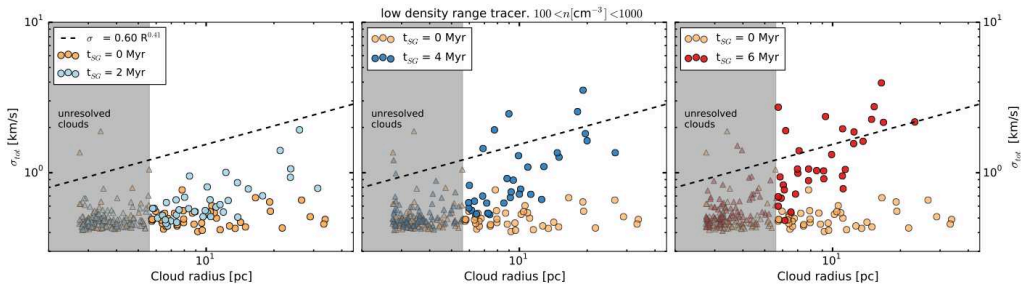} 
\caption{\label{IbanezMejia15b} Radius-velocity dispersion relation
  for clouds drawn from the models prior to the action of
  self-gravity (yellow dots in all three panels), and for clouds at
  times of 2 Myr (left), 4 Myr (center) and 6 Myr (right) after
  self-gravity begins to act.  For comparison, the observed relation
  derived from the data presented by \cite[Heyer et al.\
  (2009)]{Heyer09} is shown by the dashed line. The shaded region
  shows clouds under 10 zones whose velocity dispersion is likely to be strongly suppressed by
  numerical viscosity.  Adapted from
  \cite[Ib\'a\~nez-Mej\'{\i}a et al.\ (2015)]{IbanezMejia15}.} 
\end{center}
\end{figure}

In order to understand this rapid collapse and fragmentation, it is
necessary to examine the dynamics of these dense
regions. \cite[Ib\'a\~nez-Mej\'{\i}a et al.\ (2015)]{IbanezMejia15}
identified clouds using density contours and demonstrated that these
clouds quite uniformly had velocity dispersions $\sigma <
1$~km~s$^{-1}$, even at radii $R > 30$~pc, quite different from the
\cite[Larson (1981)]{Larson81} size-velocity dispersion relation. This
is shown by the yellow dots in all three panels of
Figure~\ref{IbanezMejia15b}, in contrast to the observed relation
presented as the dashed line, which is a fit to the subset of the
Galactic Ring Survey (\cite[Jackson et al.\ 2006]{Jackson06})
presented in \cite[Heyer et al.\ (2009)]{Heyer09}.  These low velocity
dispersions cannot resist gravitational collapse, and indeed the
clouds do collapse, forming the same sort of filamentary structures
familiar from the free gravitational collapse of the cosmic web.
Although magnetic fields are present in the simulation, they are
unable to effectively resist this collapse.  As they collapse, their
observed velocity dispersions increase to the observed values.

\cite[Ballesteros-Paredes et al.\ (2011)]{BallesterosParedes11}
emphasize that the relationship between surface density, velocity
dispersion, and radius expected from
freely falling material differs by only a factor of $\sqrt{2}$ from
that of material in virial equilibrium, and both appear consistent
with the observations as summarized by \cite[Heyer et
al. (2009)]{Heyer09}. 

It is worth noting that in a periodic box simulation of
supernova-driven turbulence, neglecting stratification, but run at
higher resolution, \cite[Padoan et al.\ (2015)]{Padoan15} find
velocity dispersions after 6~Myr that agree with our results.  However
Padoan (priv.\ comm., 2015) finds similar velocity dispersions in
their model even prior to the onset of self-gravity, in contradiction
to our results.  Further investigation will be needed to establish if
the difference is due to their higher resolution or their neglect of
stratification and confinement within the periodic box.

\section{Summary}

In this review, I have addressed a few current topics in the theory of
the atomic and molecular interstellar medium.  I have focused on the
following questions, with my current best guesses as to the answers. 
\begin{itemize} 
  \item Do cold streams contribute to gas accretion onto
  galaxies only at early times and for small galaxies, or far more
  generally as was originally claimed?  The former appears to be the
  result of the numerical simulations most likely to be reliable.
  \item How small is the actual filling factor of dense, molecular gas in
    an interstellar medium when it is observationally determined to be
    completely molecular?  Quite small, with the rest of the volume
    occupied by low-density, hot gas with temperatures exceeding $10^6$~K.
  \item Does molecular hydrogen drive star formation, or rather does
    star formation produce molecular hydrogen and other high density
    tracers as a by-product?  Molecular hydrogen and other molecules
    appear to function far more as a tracer than a cause of star
    formation, particularly since atomic line cooling is nearly as
    effective as molecular line cooling.
  \item Are the observed velocity dispersions in molecular clouds
    caused by supernova-driven turbulence, or by gravitational
    collapse?  Gravitational collapse seems more likely, but this
    question does not yet appear to be settled.
\end{itemize}

\end{document}